\documentclass[11pt,a4paper]{article}
\usepackage[utf8]{inputenc}
\usepackage[T1]{fontenc}
\usepackage{lmodern}
\usepackage{microtype}
\usepackage[
  top=1.5cm,
  bottom=2cm,
  left=1.5cm,
  right=1.5cm
]{geometry}
\usepackage{setspace}
\usepackage{parskip}
\usepackage{graphicx}
\usepackage{wrapfig}
\usepackage{placeins}
\usepackage{booktabs}
\usepackage{enumitem}
\usepackage{hyperref}
\hypersetup{colorlinks=true,linkcolor=blue,citecolor=blue,urlcolor=blue}
\usepackage{titling}
\usepackage{fancyhdr}
\usepackage{caption}
\usepackage{amsmath,amssymb}
\usepackage{titlesec}
\usepackage[
  backend=biber,
  style=numeric,
  sorting=none,
  giveninits=true,
  maxnames=1
]{biblatex}

\defbibheading{bibliography}[\refname]{\section*{#1}\vspace{-8pt}}

\titleformat{\section}
  {\large\normalfont\bfseries}
  {\thesection}{1em}{}
  
\renewbibmacro*{in:}{}

\AtEveryBibitem{%
  \clearfield{title}%
  \clearfield{subtitle}%
  \clearfield{number}%
  \clearfield{issue}%
  \clearfield{month}%
  \clearfield{day}%
  \clearfield{doi}%
  \clearfield{url}%
  \clearfield{isbn}%
  \clearfield{issn}%
  \clearfield{note}%
  \clearfield{addendum}%
  \clearfield{howpublished}%
  \clearfield{eprintclass}% kills [gr-qc]
}

\renewbibmacro*{date}{\printfield{year}}
\DeclareFieldFormat{eprint}{arXiv:#1}
\DeclareBibliographyDriver{article}{%
  \printnames{author}\setunit{\addspace}%
  \printfield{year}\setunit{\addcomma\space}%
  \printfield{journaltitle}\setunit{\addcomma\space}%
  \printfield{volume}\setunit{\addcomma\space}%
\iffieldundef{journaltitle}
  {\iffieldundef{eprint}{}{%
     \setunit{\addcomma\space}\printfield{eprint}}}
  {}%
\finentry
}
\DeclareBibliographyDriver{online}{%
  \printnames{author}\setunit{\addspace}%
  \printfield{year}%
  \iffieldundef{eprint}{}{%
    \setunit{\addcomma\space}\printfield{eprint}}%
  \finentry
}
\DeclareBibliographyDriver{unpublished}{\usebibmacro{bibindex}\usebibmacro{begentry}\usebibmacro{author/translator+others}\setunit{\addspace}\printfield{year}\iffieldundef{eprint}{}{\setunit{\addcomma\space}\printfield{eprint}}\usebibmacro{finentry}}

\defbibenvironment{bibliography}
  {}
  {}
  {\addspace
   \printtext[labelnumberwidth]{%
     \printfield{prefixnumber}%
     \printfield{labelnumber}}%
   \addhighpenspace}

\bibliography{refs}
\pretitle{\vspace*{-1cm}\begin{center}\LARGE\bfseries}
\posttitle{\par\end{center}\vskip 0.5em}
\preauthor{\begin{center}\large}
\postauthor{\end{center}}
\predate{\begin{center}\large}
\postdate{\end{center}}

\begin{document}

% -----------------------
% Cover page
% -----------------------
\begin{titlepage}
  \centering
  \vspace*{-1.7cm}
  {\Huge\bfseries Expanding Horizons \\[6pt] \Large Transforming Astronomy in the 2040s \par}
  \vspace{0cm}

  {\LARGE \textbf{Multi-messenger and time-domain astronomy in the 2040s}\par}
  \vspace{0.2cm}

  % Metadata block
  \begin{tabular}{p{4.5cm}p{12.7cm}}
    \textbf{Scientific Categories:} & Multi-Messenger Astronomy; Black Holes; Neutron Stars, Gravitational Waves; Particle Acceleration; Cosmology; Time-domain; Instrumentation \vspace{0.1cm} \\
    \textbf{Submitting Authors:} & Samaya Nissanke (DESY, German Centre for Astrophysics - DZA, University of Potsdam, DE; University of Amsterdam, NL) \\
    & Email: samaya.nissanke@desy.de \vspace{0.1cm} \\ 
    
    & Nikhil Sarin  (University of Cambridge, UK) \\
    &  Email: nikhil.sarin@ast.cam.ac.uk \vspace{0.1cm} \\
    
    & Chris Copperwheat (Liverpool John Moores University, UK ) \\
    &  Email: c.m.copperwheat@ljmu.ac.uk  \vspace{0.1cm}\\
    
    \textbf{Contributing authors:} &
    {\small
    Sarah Antier (IJCLAB, FR),
    David Berge (DESY, HU, DE),
    Pablo Bosch (UU, NL),
    Archisman Ghosh (UGent, BE),
    Paul Groot (RU, NL; UCT/SAAO, ZA),
    Gregg Hallinan (Caltech, USA),
    Tanja Hinderer (UU, NL),
    Kenta Hotokezaka (UT, JP),
    Theophanes Karydas (UvA, NL),
    Mansi Kasliwal (Caltech, USA),
    Yves Kini (UvA, NL), 
    Rubina Kotak (UTU, FI),
    Kumiko Kotera (IAP, FR),
    Marek Kowalski (DESY, HU, DE), 
    Luke Krauth (UvA, NL),
    Kruthi Krishna (DESY, DZA, DE),
    Thomas Kupfer (UHH, DE),
    Paraskevas Lampropoulos (UvA, NL),
    Andrew Levan (RU, NL),
    Ioannis Liodis (DESY, DZA, DE),
    Lea Marcotulli (DESY, DE), 
    Kunal Mooley (IIT, IN),
    Silvia Piranomonte (INAF, IT), 
    Nanda Rea (ICE-CSIC, ESP), 
    Martin Roth (AIP, DZA, DE),
    Simone Scaringi (Durham, UK), 
    Steve Schulze (U Northwestern, USA),
    Lami Suleiman (DESY, DZA, DE),
    Nial Tanvir (Leicester, UK),
    Angela Zegarelli (RUB, DE), 
    Sylvia J. Zhu (DESY, DE)
    }\vspace{0.1cm}\\
    
    \textbf{Endorsers:}  & 
    {\small
    Shin'ichiro Ando (UvA, NL),
    Igor Andreoni (UNC, USA),
    Iair Arcavi (TAU, IL),
    Smaranika Banerjee (TU Darmstadt, DE),
    Sudhanshu Barway (IIA, IN),
    Maria Grazia Bernardini (INAF, IT),
    Gianfranco Bertone (UvA, NL),
    Varun Bhalerao (IITB, IN),
    Stéphane Blondin (ESO, DE; LAM, FR),
    David Buckley (SAAO/UCT, ZA),
    Mattia Bulla (UniFE, IT),
    Jonathan Carney (UNC, USA),
    S.~Bradley Cenko (NASA/GSFC, USA),
    Ashley Chrimes (ESA/ESTEC, NL),
    Teagan Clarke (Princeton, USA),
    Deanne L. Coppejans (Warwick, UK),
    Michael W. Coughlin (UMN, USA),
    Paolo D'Avanzo (INAF, IT),
    Benjamin L. Davis (NYUAD, UAE),
    Valerio D'Elia (ASI, IT),
    Tim Dietrich (UP, AEI, DE), 
    Dougal Dobie (USYD, AU),
    Nancy Elias-Rosa (INAF, IT),
    Francois Foucart (UNH, USA),
    Anna Franckowiak (RUB, DE),
    Morgan Fraser (UCD, IE),
    Matthew J. Graham (Caltech, USA),
    Alexander Heger (Monash, AU),
    Mich\`{e}le Heurs (LUH, DZA, DESY, DE),
    Andreas Haungs (KIT, DE),
    Luca Izzo (INAF, IT),
    Valeriya Korol (MPA, DE), 
    Hanindyo Kuncarayakti (UTU, FI),
    Gavin P Lamb (LJMU, UK),
    Paul Lasky (Monash, AUS),
    Sera Markoff (UvA, NL; CAM, UK),
    Antonio Martin-Carrillo (UCD, IRE),
    Seppo Mattila (UTU, Fi),
    Andrea Melandri (INAF, IT),
    Anais Möller (SUT,AU),
    Bernhard Müller (Monash, AU),
    Thomas Murach (DESY, DE),
    Anna Nelles (FAU, DESY, DE),
    Kieran O'Brien (Durham, UK),
    Paul O'Brien (UoL, UK),
    Antonella Palmese (CMU, USA; INAF, IT),
    Kerry Paterson (MPIA, DE),
    Quentin Pognan (AEI, DE),
    Rafael A. Porto (DESY, DE),
    Liana Rauf (ANU, AU),
    Andrea Sanna (UniCA, IT),
    Avinash Singh (OKC, SU, SE),
    Danny Steeghs (Warwick, UK),
    Christian Stegmann (DESY, DZA, UP, DE),
    Leo Stein (Mississippi, USA),
    Robert Stein (UMD, USA),
    Gianpiero Tagliaferri (INAF, IT),
    Aishwarya Linesh Thakur (INAF, IT),
    Silvia Toonen (UvA, NL),
    Antonio de Ugarte Postigo (LAM, FR),
    Natasha Van Bemmel (SUT, AU),
    Anna Watts (UvA, NL),
    Helvi Witek (UIUC, USA)
    }
  \end{tabular}

\end{titlepage}
\section{Introduction and Background}
\label{sec:intro}
\vspace{-8pt}
Once largely theoretical, multi-messenger astronomy (MMA) is now an observational discipline. The detection of neutrinos from SN1987A confirmed core-collapse and proto–neutron star cooling, revealing physics inaccessible to electromagnetic (EM) observations \cite{Hirata87}. The binary neutron star merger GW170817, the first gravitational wave (GW) source with identified EM counterparts \cite{Abbott2017a,Abbott2017b}, demonstrated the power of joint observations: GWs directly  measured masses and distance, whereas EM data provided  the location and redshift, revealed a structured relativistic jet powering the associated GRB 170817A, identified a lanthanide-rich kilonova (AT2017gfo), and enabled an independent measurement of the Hubble constant. Despite progress, MMA discoveries remain rare: GW170817 is still the only confirmed GW-EM event among hundreds of GW detections \cite{2025arXiv250818082T}. Most GW mergers are localised from 100s to 1000s of square degrees at distances of hundreds of Mpc to Gpcs, severely limiting follow-up, an issue compounded by the low binary neutron star (BNS) merger rate. 

Regarding high-energy neutrinos, exciting evidence for sources such as the Blazar TXS 0506+056 \cite{txs,txs2} are emerging. Whereas these demonstrate the importance of high-energy neutrinos for understanding hadronic cosmic ray accelerators, more associations are required to obtain a mature picture of the processes involved.

\begin{wrapfigure}{l}{0.55\textwidth}
\vspace{-10pt}
  \centering
  \begin{minipage}{0.55\textwidth}
    \centering
    \includegraphics[width=\linewidth]{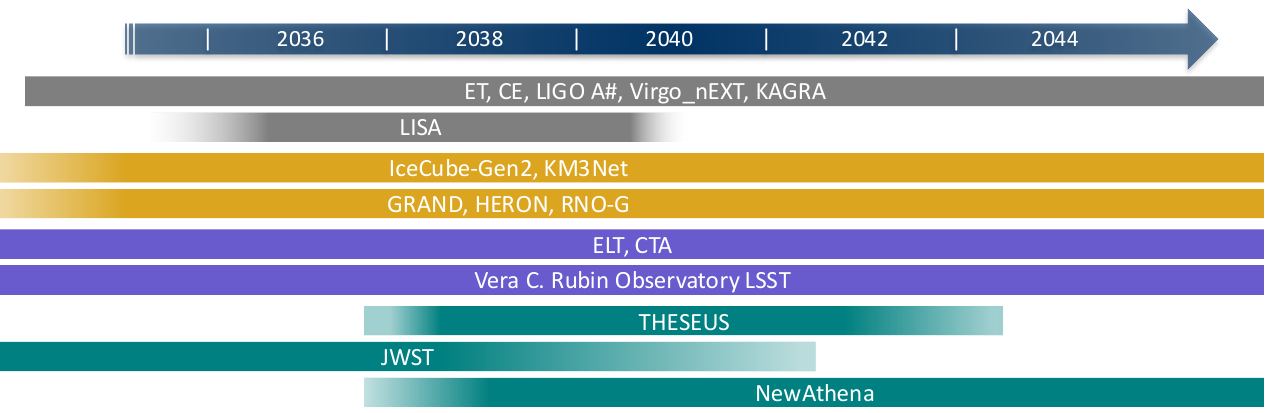}
  \end{minipage}\hfill
\caption{\small Timelines for a subset of MMA facilities in the 2040s.}
\label{figure1}
\vspace{-15pt}
\end{wrapfigure}

The coming two decades will deliver a suite of transformative facilities that will define the MMA landscape of the 2040s (Fig.~\ref{figure1}). Next generation ground-based GW observatories, such as the Einstein Telescope \cite{Abac2025} and Cosmic Explorer \cite{evans2021horizonstudycosmicexplorer}, are expected to detect $\sim \mathcal{O}(10^{5})$ BNS mergers yr$^{-1}$ out to $z$ of a few, with hundreds yr$^{-1}$ localised to a hundred square degrees ten minutes \emph{prior} to coalescence, together with $\mathcal{O}(10^{4}$–$10^{5})$ neutron star–black hole (NSBH) mergers yr$^{-1}$, $\mathcal{O}(10^{5})$ binary black hole (BBH) mergers yr$^{-1}$ with some fraction expected to be embedded in the disks of Active Galactic Nuclei (AGN), and potentially signals from Galactic neutron stars (NSs) and nearby core-collapse supernovae (CCSNe). At lower frequencies, space-based GW missions, notably \textit{LISA}, will provide weeks-to-months early warning of supermassive BBH mergers, extreme mass-ratio inspirals, and compact Galactic binaries, enabling coordinated follow-up \cite{LISA:2024hlh}. Pulsar timing arrays are extending GW observations to nano-Hz frequencies, probing the population of gradually inspiralling supermassive BBHs across cosmic time and will enable single source detections. Next generation neutrino observatories such as IceCube-Gen2 \cite{2021JPhG...48f0501A} and KM3NeT \cite{km3net} will improve high-energy sensitivity by an order of magnitude, and deliver hundreds of cosmic neutrino alerts per year, each pointing to a source with $\sim$~degree resolution and thus enabling routine identification of astrophysical neutrino sources. At ultra-high energies, observatories such as GRAND \cite{álvarezmuñiz2025giantradioarrayneutrino}, HERON \cite{kotera2025hybridelevatedradioobservatory}, and RNO-G \cite{agarwal2025instrumentdesignperformanceseven} will target and identify rare neutrino events with deep instantaneous sensitivity and sub-degree angular resolution \cite{kotera2025observationalstrategiesultrahighenergyneutrinos}. Operated in concert, $\gamma$-ray facilities such as CTAO \cite{CTAOScience} and future MeV missions will probe the extreme particle acceleration environments that also produce GW and neutrino emission. %They will complement new facilities recently in operation with as SVOM or Einstein Probe, providing regular GRB triggers for the next decade. 
Parallel advances across the EM spectrum, including the Vera C. Rubin Observatory LSST, the ESO’s ELT and VLT, SKA, and future UV and X-ray missions, will deliver redshifts and detailed environmental and host galaxy characterisation of compact object mergers across cosmic time. \textbf{Together, this ensemble of facilities will generate} 
\textbf{an unprecedented volume of MMA triggers, demanding a dedicated large aperture time-domain facility with comprehensive northern and southern hemisphere coverage to provide the rapid, sensitive, and spectroscopically flexible EM characterisation required to fully exploit this emerging MMA ecosystem.}

\vspace{-10pt}
\section{Open Science Questions}
\label{sec:openquestions}
\vspace{-8pt}
Below we outline a non exhaustive set of key science themes that MMA will address in the 2040s.

\textbf{2.1 \, Chemical Evolution and Nucleosynthesis}: 
Samples of $\mathcal{O}(10^{5})$ or more of BNS and NSBH events (or more exotic objects such as sub-solar NS/BHs) will constrain uncertain GW merger rates. Combined with abundance estimates from UVOIR kilonova spectra (such as Sr II, Ce III, Te III spectral lines in the near infrared (nIR)) of tens to hundreds of such events, these observations will trace the production of r-process elements across cosmic time, quantify the fraction produced by compact mergers, and reveal its dependence on merger properties, such as mass ratios, tidal deformations, and BH spins. Achieving this goal requires rapid response, deep and multi-epoch spectroscopy. \\
\textbf{2.2 \, Jet Physics}: GWs will provide independent measurements of distance and inclination, constraining viewing angles to the $\sim \%$-level for high signal-to-noise mergers. EM observations from the optical--nIR to radio and TeV energies quantify jet angular structure, Lorentz factors, magnetisation, dissipation radii, and any thermal emission through time-resolved spectroscopy, lightcurve evolution, polarimetry, and high-resolution imaging. Together, these measurements will break degeneracies between geometry, microphysics, and environment, enabling tests of jet launching mechanisms from NSs and/or BHs. \\
\textbf{2.3 \, Cosmic Particle Acceleration}:
Coincident observations of high-energy neutrinos and EM emission will pinpoint cosmic-ray accelerators, separating hadronic from leptonic channels. It will enable unique insights into the extreme particle acceleration mechanism, required to explain the observed high-energy cosmic rays and neutrinos.  In connection with Sec.\ 2.1, this will allow the study of the role of relativistic jets in high-energy particle production and cosmic feedback over cosmic history.\\
\textbf{2.4 \, Stellar Evolution}:
Rates of compact object mergers (BNS, NSBH, BBH) combined with EM properties of host galaxies constrain delay times and evolutionary paths. For rare events, e.g. Galactic SNe, GWs provide information on asymmetries during core collapse, neutrino light curves probe core conditions, and EM data reveal ejecta composition and how the shock progressed through the star.\\
\textbf{2.5 \, Supermassive Black Holes and their growth}: As LISA will localise merging supermassive black hole (SMBH) binaries to tens or thousands of square degrees a week before merger \cite{LISA:2024hlh}, early identification and redshift determination of EM counterparts is crucial for coordinated observation campaigns. Searches for variability patterns in AGN disks (sampled by spectroscopic campaigns) and jets (sampled by CTAO) caused by SMBH binary interactions can uncover PTA counterparts and the progenitors of LISA sources \cite{2009ApJ...697.1621G,2024MNRAS.52710168P}. EM detections reveal the interactions between the gas, magnetic fields, and SMBHs in galaxies at late stages in their binary evolution, which are not well understood. Additionally, merging SMBH binaries could contribute to the diffuse astrophysical neutrino flux \cite{2023MNRAS.518.6158J} and produce transient neutrino emission (e.g., \cite{Britzen+TXS}).
Large localisations, variety of timescales and uncertainty on the nature of the EM counterparts necessitate a large aperture EM observatory with flexible capabilities.\\
\textbf{2.6 \, Cosmology and Fundamental Physics}: MMA observations provide a powerful means of addressing key cosmological questions, including the nature of dark energy and the possibility of
\begin{wrapfigure}{r}{0.52\textwidth}
    \vspace{-13pt}
    \centering
    \includegraphics[trim={0 10pt 0 0},clip,width=\linewidth]{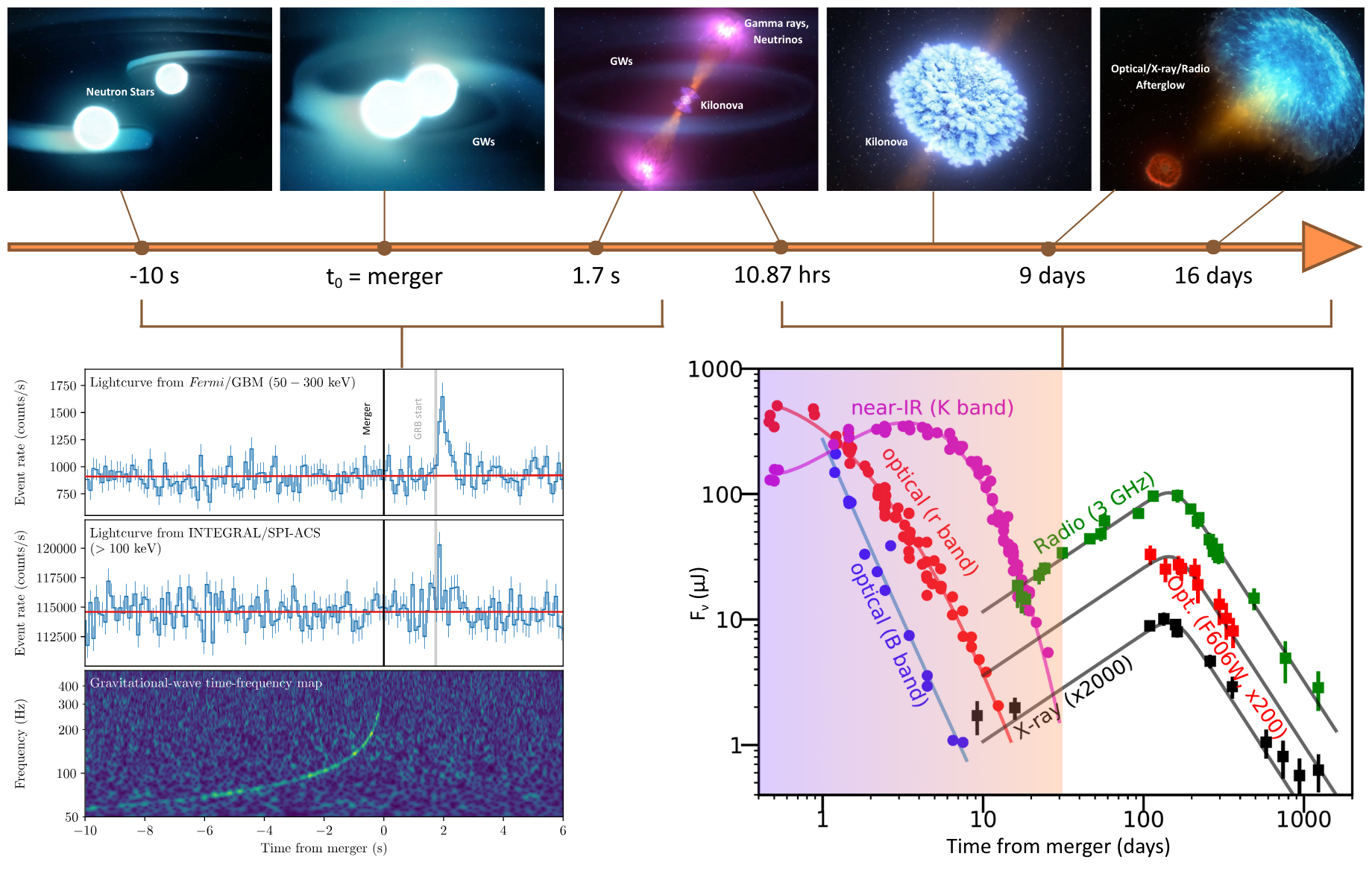}
    \caption{\small Top panel shows the evolution timescales of GW170817 and its counterpart, with detections across a selection of wavebands indicated. The bottom left panel illustrates the GW chirp and associated high-energy prompt emission. The bottom right panel shows the evolution of the spectral flux density over the first several hundred days of the kilonova and GRB afterglow. Rapid changes in brightness and spectral properties during the first several days (shaded region) highlight the necessity of intensive, rapid-response follow-up observations from the UV to the NIR, which are also a prerequisite for successful counterpart identification. (Figure adapted from NASA; \cite{Abbott2017b}; Kunal Mooley).}
    \label{figure2}
    \vspace{-20pt}
\end{wrapfigure}
modified gravity on cosmological scales through their distinct propagation effects. One of the hallmark results from the BNS merger GW170817 was the first MMA bright standard siren measurement of the Hubble constant $H_0$ \cite{Abbott2017c}. In the 2040s, bright standard sirens are expected to measure $H_0$ to sub-percent precision \cite{Gupta_2024}, and to probe late time cosmological parameters, particularly the dark energy equation of state. Furthermore, energy dependent neutrino arrival times will test Lorentz invariance at exceptional sensitivity. Together, MMA observations of compact object mergers and SNe will enable stringent tests of fundamental physics.\\
\textbf{2.7 \, Dense Matter Equation of State}: BNS and NSBH mergers probe the high-density, low-temperature regime that complements the QCD conditions studied in terrestrial heavy-ion collision experiments. Modelling the equation of state (EoS) in this regime is challenging due to strong interaction uncertainties at high densities. Matter may include exotic components -- hyperons, pion condensates, or deconfined quark matter -- in its different superfluid states, beyond the standard npe$\mu$ composition. These variations lead to wide differences in NS properties such as maximum mass, compactness, and tidal deformability, rendering its signatures degenerate; making it challenging to constrain accurately the EoS and the NS's interior structure. Determining the EoS of dense matter is essential to understand NS properties across diverse environments and dynamical scenarios leading to multimessenger emission. The discovery of GW170817 and its EM counterparts excluded overly stiff EoS that predict large tidal deformabilities (e.g., \cite{Raaijmakers_2020,2020Sci...370.1450D}). In the 2040s, MMA signals will probe with exquisite precision the EoS \cite{Gupta_2024}, mapping out any first-order phase transitions from hadronic to dense quark matter.
\vspace{-10pt}
\section{Technology Requirements}
\label{sec:tech}
\vspace{-8pt}

Addressing the science cases outlined above requires a coordinated, multi-tiered, multiwavelength follow up strategy (Fig.~\ref{figure2}) with advances spanning theory, analysis and instrumentation. Ground-based capabilities must include large aperture telescopes capable of identifying candidate counterparts on hour timescales when no high-energy or radio localisation is available, with O–NIR spectroscopy providing definitive kilonova confirmation and physical characterisation. In the 2040s, astroparticle facilities will deliver a dramatically increased rate of triggers, the majority with large sky localisations and at greater distances than today, demanding follow-up capabilities in both hemispheres and across a wide range of longitudes. A critical capability gap is rapid, flexible O–nIR spectroscopic response. Spectroscopy in these bands can  determine ejecta composition, and provide redshift measures, energetics, and viewing geometry, information inaccessible to GWs, neutrinos, $\gamma$-ray or radio observations alone. A large-aperture facility with low-latency response, efficient target acquisition, and high-throughput spectroscopy is essential; however, we do not currently expect a flexible, dedicated facility of this type to be available by the 2030s. Without this, the 2040s time-domain landscape faces a major bottleneck, limiting the science return from the unparalleled volume and diversity of transients delivered by next generation facilities.\\
\textbf{Requirements}: (i) a telescope dedicated to time domain science, since the demand for target-of-opportunity observations would overwhelm and disrupt the broader observing programs on classically and queue scheduled telescopes; (ii) effective collecting area equivalent to at least a $20$--$30$ metre class telescope for $R \sim 2\mathord,000 - 10\mathord,000$ spectroscopy at $m \sim 25$; (iii) dynamic scheduling for real-time prioritization and rapid ($\sim$seconds) response to capture phenomena such as shock breakouts and pre-merger precursors; (iv) flexible rapid/simultaneous response to many candidates spanning a wide range of magnitudes and evolutionary time scales, spread over the full sky; (v) a wavelength range that can monitor kilonova evolution from the early blue optical ($300-1\mathord,000\,$nm) emission to the later red optical/nIR lanthanide rich ejecta ($1\mathord,000-2\mathord,500\,$nm). 
A telescope concept that meets these requirements is one or more multi-aperture facilities composed of independently steerable telescopes feeding a common spectrograph capable of tiling large regions or combining apertures for deep spectroscopy. Low read-noise, high-throughput detector technologies such as qCMOS \cite{10.1117/12.3055930}, Skipper CCDs \cite{Tiffenberg2017}, APD arrays \cite{Bradford2025}, MKIDs \cite{Mazin13}, and photonics lanterns \cite{10.1117/12.2629901} enable this approach. A telescope of this type represents a step change in capability for MMA follow-up and is a transformational facility across the entire field of astronomy.
\vspace{-10pt}
\printbibliography

\end{document}